\documentclass[11pt,english,eqsecnum,prd,aps,superscriptaddress,longbibliography,tightenlines,nofootinbib]{revtex4-2}
\usepackage[T1]{fontenc}
\usepackage[latin9]{inputenc}
\usepackage{color}
\usepackage{babel}
\usepackage{amsmath}
\usepackage{graphicx}
\usepackage{float}
\PassOptionsToPackage{normalem}{ulem}
\usepackage{ulem}
\usepackage[unicode=true,pdfusetitle,
bookmarks=true,bookmarksnumbered=false,bookmarksopen=false,
breaklinks=false,pdfborder={0 0 1},backref=false,colorlinks=true]
{hyperref}
\hypersetup{allcolors=blue}
\usepackage{tikz}
\usetikzlibrary{decorations.pathmorphing}

\makeatother

\begin{document}
	\title{Model metrics for quantum black hole evolution: Gravitational collapse, singularity resolution, and transient horizons}
	\author{Samantha Hergott}
	\email{sherrgs@yorku.ca}
	\affiliation{Department of Physics and Astronomy, York University, 4700 Keele Street, Toronto, Ontario M3J 1P3, Canada}

	\author{Viqar Husain}
	\email{vhusain@unb.ca}
	\affiliation{Department of Mathematics and Statistics, University of New Brunswick,
		Fredericton, NB, Canada E3B 5A3}

	\author{Saeed Rastgoo}
	\email{srastgoo@yorku.ca}
	\affiliation{Department of Physics, University of Alberta, Edmonton, Alberta T6G 2G1, Canada}
	\affiliation{Department of Mathematical and Statistical Sciences, University of Alberta, Edmonton, Alberta T6G 2G1, Canada}
	\affiliation{Theoretical Physics Institute, University of Alberta, Edmonton, Alberta T6G 2G1, Canada}
	\affiliation{Department of Physics and Astronomy, York University, 4700 Keele Street, Toronto, Ontario M3J 1P3, Canada}


	\begin{abstract}
		It is widely accepted that curvature singularity resolution should
		be a feature of quantum gravity. We present a class of time-dependent asymptotically flat spherically symmetric metrics that model gravitational collapse in quantum gravity. The metrics capture intuitions associated with the dynamics of singularity resolution, and horizon formation and evaporation following a matter bounce. A parameter in the metric associated with the speed of the bounce determines black hole lifetime as a power of its mass; this includes the Hawking evaporation time $M^{3}$.
		
	\end{abstract}
	
	\maketitle
	
	\section{Introduction}
	
	There are presently several approaches to quantum gravity. These differ
	in starting assumptions, including the structures that are to remain
	fixed before quantization is implemented \citep{Isham:1995wr,Carlip:2017dtj}. Guided by the quantum resolution of the Coulomb singularity in electromagnetism, all approaches agree on the necessity of resolving spacetime curvature singularities. For cosmology this is the Big Bang singularity, and for black holes it is the singularity inside the event horizon.  
	
	To date most work has focused on cosmological singularites. The corresponding models are quantum mechanical with a few degrees of freedom \citep{Misner:1969hg,Blyth:1975is,Agullo:2016tjh} rather than field theoretic, and are therefore relatively easier to quantize than inhomogeneous models that describe black hole formation. Although there is much work on gravitational collapse in classical general relativity, there is relatively little work to date on whether and how collapsing matter bounces and dynamically avoids singularity formation in quantum gravity; a recent work on dust collapse is a notable exception \citep{Kelly:2020lec,Husain:2021ojz,Husain:2022gwp}.
	
	It is known from Hawking's work that black holes are semiclassically
	unstable \cite{Hawking:1975vcx}, and evaporate by emitting radiation of quanta of matter fields propagating on the black hole background. This is a low energy result that remains robust under modification of the dispersion relations of the field with a Planck-scale cutoff \citep{Jacobson:1999ay,Unruh:2004zk}. The evaporating black hole has led to the well-known ``information loss'' conformal diagram (pg. 219 of \cite{Hawking:1975vcx}). This picture has been the focus of much attention even though it retains the spacelike curvature singularity inside the event horizon and is not the conformal diagram of any explicit metric. (See  Ref. \cite{Mann:2021mnc} for a recent review.)
	
	This unsatisfactory feature has led to models of non-singular black
	holes from various directions, with the aim of proposing alternative
	conformal diagrams which include quantum gravity effects \citep{Ashtekar:2005cj,Hayward:2005gi,Martin-Dussaud:2019wqc}. Although these are useful for discussing possibilities, a conformal diagram must ultimately come from an actual ``effective'' metric that is derived from a theory of quantum gravity.  (See  Ref. \cite{Mann:2021mnc} for a recent review.)
	
	A related question is whether the global teleological object that
	is the event horizon should remain as a sacred classical structure
	that survives quantization, or whether it is replaced by dynamical
	horizons that form and ultimately disappear.  Recent analytical and numerical work on dust collapse with quantum gravity corrections provides evidence for the latter outcome; it suggests that black holes end with shock wave emission \cite{Husain:2021ojz}. Whether this results holds for other types of matter is an open question.
	
	Lastly, it is uncontroversial that classical gravitational collapse leads to a unique long-time static limit  which is a black hole, at least for sufficiently dense concentrations of matter. What is the corresponding situation for gravitational collapse with quantum gravity features built into models? If a black hole is no longer a universal attractor, and singularity avoidance causes a matter bounce, the details could be matter-type dependent and not universal. This is because a matter bounce would necessarily involves dynamics intrinsic to its unique Lagrangian.  
	
	Motivated by these questions, and past work on non-singular black
	holes, we propose and study a class of time dependent, asymptotically
	flat, spherically symmetric metrics that model dynamical singularity
	avoidance. As such, our work may be viewed as extending the ideas in
	\citep{Hayward:2005gi}, and motivated by the recent results on dust collapse \citep{Husain:2021ojz} mentioned above.
	
	The metrics we describe are naturally written in the Painleve-Gullstrand (PG) coordinates \citep{Guven:1999hc,Martel:2000rn} with a specific form of time-dependent mass function that describes inflow followed by outflow of matter.  PG coordinates cover the classical spacetime of matter collapse, and there are no known Kruskal-like extensions with matter fields. Therefore these coordinates should also be sufficient to capture a matter bounce. The class of metrics we describe, exhibit horizon formation and evaporation with a temporary event horizon, and provide a parameter-dependent formula for black hole lifetime.
	Although not a solution to any specific model of quantum gravity,
	the metrics smoothly capture essential features expected of singularity
	resolution and matter bounce in quantum gravity. This is similar in
	spirit to various types of ``metric engineering'', including wormholes
	\citep{Morris:1988tu} and warp drives \citep{Alcubierre:2017pqm}.
	
	\section{Modeling quantum gravitational collapse}
	
	Our goal is to construct spherically symmetric metrics that can model
	gravitational collapse and bounce in quantum gravity, which are asymptotically
	flat and represent the Schwarzschild solution outside an inner radial
	region.
	
	The generic spherically symmetric metric may be written in the form
	\begin{equation}
		ds^{2}=-N^{2}(r,t)dt^{2}+f^{2}(r,t)\left(dr+N^{r}(r,t)dt\right)^{2}+g^{2}(r,t)d\Omega^{2}\label{eq:Gen-metric}
	\end{equation}
	where $N$ is the lapse, $N^{r}$ is the shift, $r$ is a radial coordinate
	and $f(r,t),\,g(r,t)$ are arbitrary functions. A particular and useful
	special case of this metric is the Painleve-Gullstrand form \citep{Martel:2000rn,Guven:1999hc}
	obtained by setting $g(r,t)=r$, $N(r,t)=1=f(r,t)$, and 
	\begin{equation}
		N^{r}(r,t)=\sqrt{\frac{2Gm(r,t)}{r}}.
	\end{equation}
	This defines the mass function $m(r,t)$. This function gives the
	mass contained within a radius $r$ at a time $t$ and is related to
	the density by 
	\begin{equation}
		m(r,t)=4\pi\int_{0}^{r}r^{2}\rho(r,t)dr.
	\end{equation}
	If $\rho(r,t)>0$, then $m(r,t)$ is an increasing function with the
	property that 
	\begin{equation}
		\label{ADM}
		\lim_{r\to\infty}m(r,t)=M
	\end{equation}
	for asymptotically flat metrics. $M$ is then the  Arnowitt-Deser-Misner
	mass. If $r^2\rho(r,t)$ goes to zero near the origin $r=0$, $m(r,t)$
	rises from zero and asymptotes to the value $M$; the corresponding
	density $\rho(r,t)$ is non-zero in the regions where the mass function
	increases monotonically with $r$.
	
	These considerations suggest modeling collapse and bounce by using
	mass functions that initially move toward $r=0$, reverse direction
	at some small $r$, and then move outward. While far from being unique,
	these features are efficiently captured by the function 
	\begin{equation}
		m(r,t)=M_{0}\left[1+\tanh\left(\frac{r-r_{0}-v(r,t)t}{\alpha\  l_{0}}\right)\right]^{a}\tanh\left(\frac{r^{b}}{l_{0}^{b}}\right).\label{m}
	\end{equation}
	Here $l_{0}$ is a scale (typically the Planck length); $r_{0}$ is associated to the radial position of the bounce (as we will see below); $M_{0}$ scales the total ADM mass $M$; $\alpha$ determines how sharply the mass function rises; 
	$a$ and $b$ are positive integers (or reals) to be chosen such that
	the density 
	\begin{equation}
		\rho(r,t)=\frac{1}{4\pi r^{2}}\frac{\partial m}{\partial r}\label{rho}
	\end{equation}
	is bounded as $r\rightarrow0$, a feature necessary for singularity
	avoidance; $v(r,t)$ is the velocity function of the density profile. The hyperbolic tangent function is the simplest choice
	that captures the desired properties of the mass function.
	
	The velocity function is to be chosen such that an initially inward
	falling mass profile bounces and starts to move outward, and 
	the profile speeds up as it falls inward and slows down as it moves
	outward. These features are captured in the function 
	\begin{equation}
		v(r,t)=\frac{A}{(1+r)^{n}}\tanh\left(\frac{t-t_{0}}{l_{0}}\right),\label{v}
	\end{equation}
	where $A>0$ is a real constant $t_{0}$ is the time when the bounce occurs, and $n$ is a positive constant (which, as we will see, determines the black hole lifetime).
	
	This completes our description of the metric designed to model non-singular
	gravitational collapse and bounce, as might emerge from a quantum theory
	of gravity. As we have mentioned, such metrics are not unique, but the one we present captures interesting dynamical features such as ``horizon bubbles'' speculated on in \citep{Hayward:2005gi}.
	
	\section{Physical features}
	
	We describe several interesting properties that can be extracted from
	the above class of metrics, and indeed many others like it. These are the evolution of mass and density,
	horizon formation and evaporation, and black hole lifetime. The last
	feature is a function of the power $n$ in the velocity function \eqref{v} defined
	above.
	
\begin{figure}[H]
		\includegraphics[width=\columnwidth]{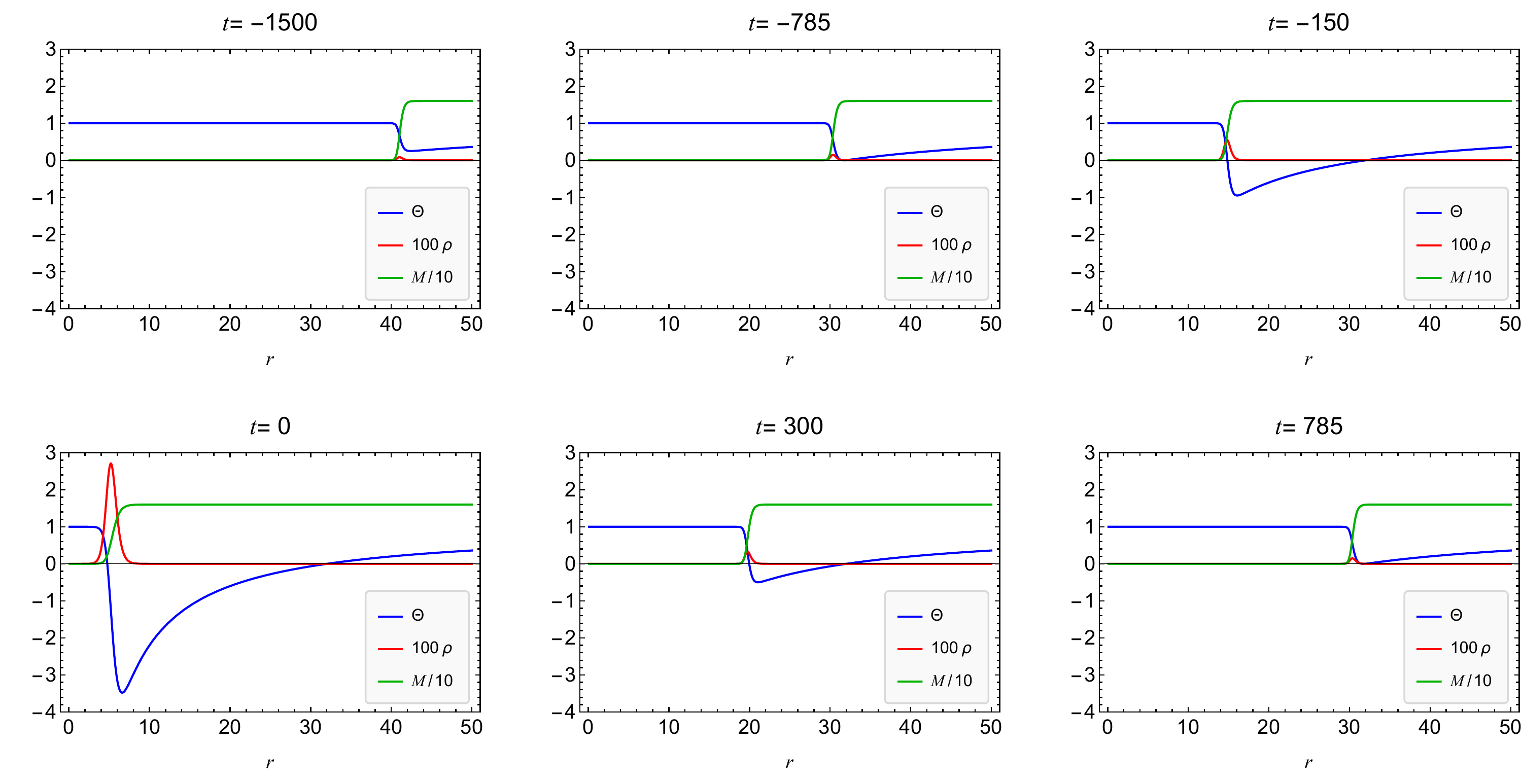}
		\caption{Time sequences of the mass function $m$ (green), density $\rho$ (red), and horizon function $\Theta$ (blue): the first row shows in-going matter with horizons (roots of the $\Theta$) appearing, and the second row shows matter bouncing and moving outward. Horizons eventually disappear by the last frame. The outer horizon for most of the evolution remains at a fixed location until it disappears with the root of $\Theta$. \label{fig:Mass-Density-grr}}
	\end{figure}

	Fig. \ref{fig:Mass-Density-grr} shows time sequences of the mass function \eqref{m},
	density \eqref{rho}, and the apparent horizon function 
	\begin{equation}
		\Theta(r,t)=g^{ab}\partial_{a}r\partial_{b}r=g^{rr}=1-(N^r)^2
	\end{equation}
	whose roots give the horizon trajectories $r_{H}(t)$. For the metric parametrization (\ref{eq:Gen-metric}), this coincides with the expansion determined by the Lie derivative ${\cal L}_{n_{+}} \omega = (1-N^r)\ \omega$ of the 2-sphere area form
	$\omega = r^2 \sin\theta\ d\theta\wedge d\phi$ with respect to the future pointing outgoing radial null vector $n_+^a = (1,1-N^r$). 
	\begin{figure}[H]
		\includegraphics[width=\columnwidth]{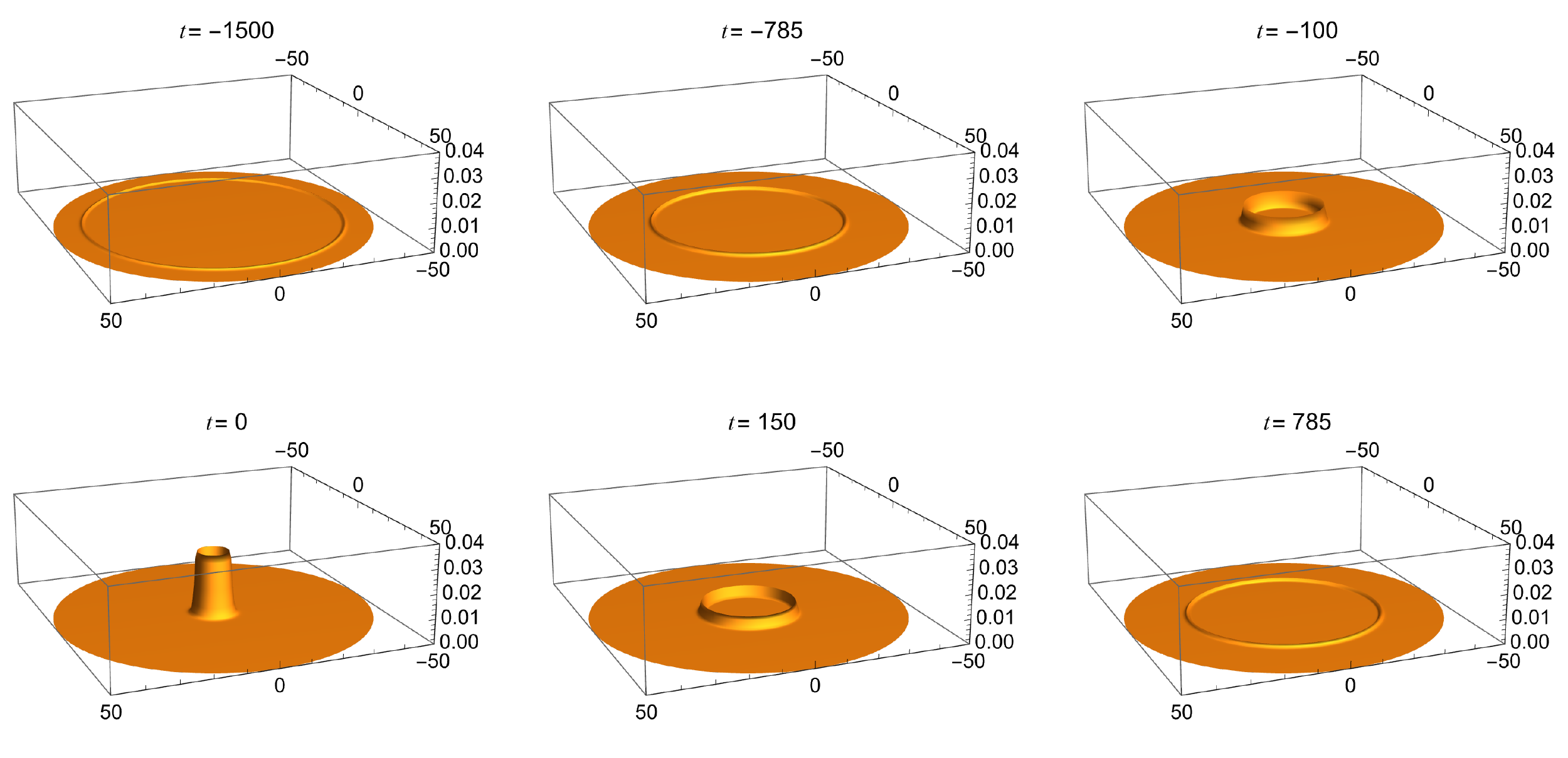}
		\caption{Density function $\rho(r,t)$ at the indicated times. The frames top-left to bottom-right show how the ingoing density profile nears $r=0$, bounces, and moves outward.}
		\label{density-3d}
	\end{figure}
	
	These quantities are scaled in Fig. \ref{fig:Mass-Density-grr} to fit in the same frames. In this figure, the first row shows
	the incoming density and horizons (the two roots of $\Theta$) appearing, and
	the second row shows matter bouncing and horizons disappearing as
	matter moves outward. Fig. \ref{density-3d} shows two-dimensional
	bouncing density profile. The parameters used for Figs. 1 to 5, and 7, are $a=2$, $b=3$, $\alpha=1$, $M_0=4$, and $r_0=5$ in the mass function \eqref{m}, and $A=n=1$ in the velocity function (\ref{v}), (except for Fig. 7 which exhibits three values of $n$). These values are chosen for illustration; qualitatively similar  results arise  for other choices.

	\subsection{Microscopics of horizons}
	
\begin{figure}
		\begin{centering}
			\includegraphics[scale=0.6]{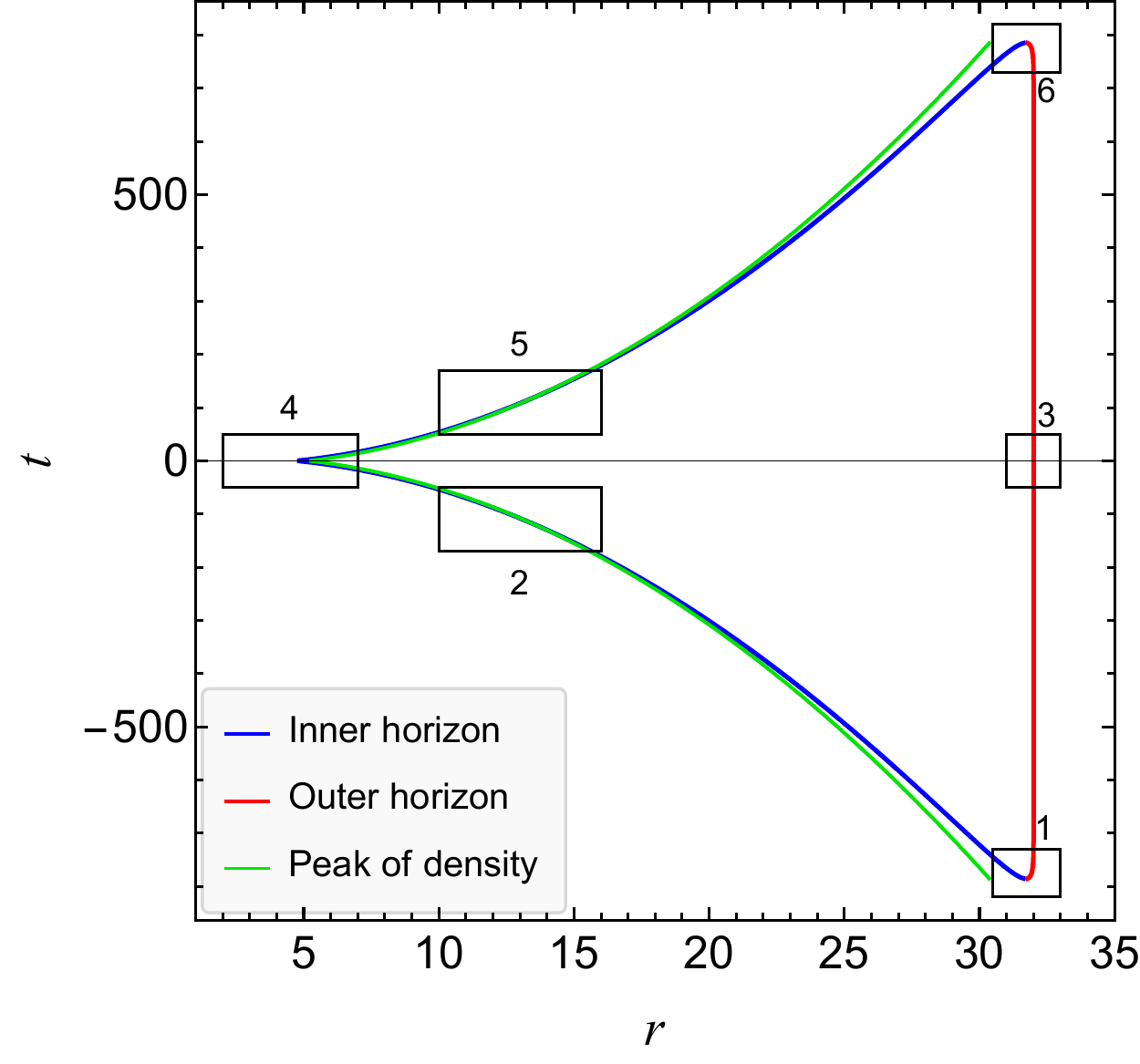}
			\par\end{centering}
		\caption{Horizon and density peak trajectories. Horizons form and diverge from point 1, density peak (green) and inner horizon (blue) bounce at point 4, and the inner (blue) and outer (red) horizons merge and annihilate at point 6 (details of each region appear magnified in the next figure). \label{fig:HorizonsCausal-all}}
	\end{figure}

\begin{figure}
		\includegraphics[width=\columnwidth]{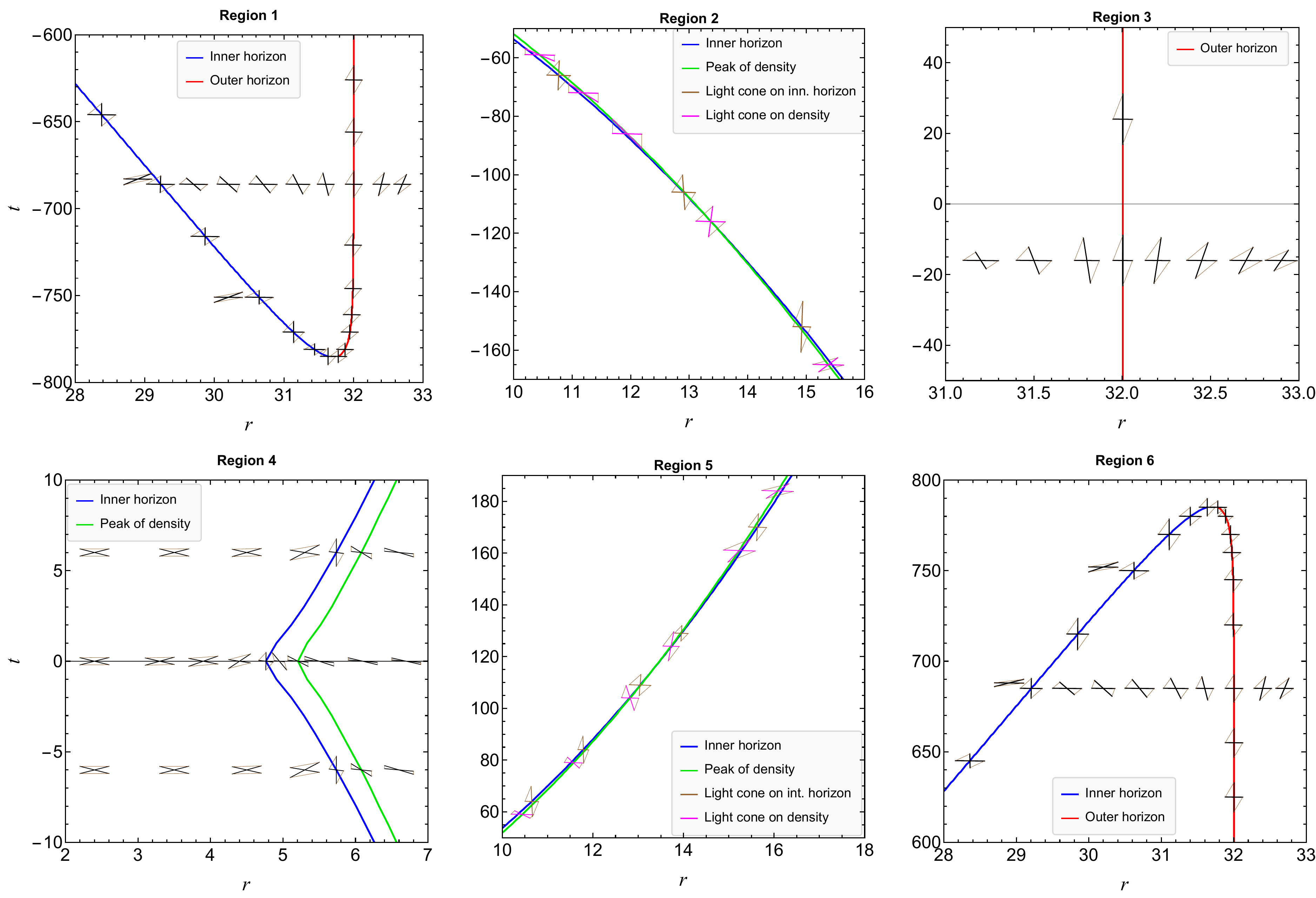}
		\caption{ Close up of regions 1-6 of Fig. \ref{fig:HorizonsCausal-all}. R1: Horizons form. The outer one becomes null and stays at constant radius, while the inner one is timelike. R2: the peak density crosses from the interior of the inner horizon to the region between the two horizons. R3: the outer horizon becomes null after its formation. R4: the inner horizon and matter density bounce. The inner horizon goes from timelike to spacelike and the peak matter density stays spacelike inside the region between the two horizons. Lightcones near $r=0$ resemble those of flat spacetime. R5: matter density peak emerges from within the region between the two horizons and becomes timelike. R6: inner and outer horizons merge and annihilate.} 
		\label{fig:HorizonsCausal-all-2}
	\end{figure}

	A detailed look at the dynamical horizons with the mass function used
	for the first two figures reveals interesting features of the trajectories of the horizons and peak value of the density.   Fig. \ref{fig:HorizonsCausal-all} shows the horizon trajectory in $r,t$ plane: the red line is the outer horizon, the blue one is the inner horizon, and the green line shows the peak of
	the density function. A pair of horizons form in region 1 and begin to diverge. The inner horizon and density lines then move inward together, bounce in region 4, and then travel outward until the inner
	and outer horizons merge and annihilate in region 6. The outer horizon (red) is null for almost the entirety of its lifetime. 
	\noindent Details of each boxed region in Fig. \ref{fig:HorizonsCausal-all} appear in Fig. \ref{fig:HorizonsCausal-all-2}, together with the
	local light cones at selected points; all figures and light cones therein are from direct calculations, except the conformal diagram (Fig. \ref{fig:PenroseD}), which is inferred from Fig. \ref{fig:HorizonsCausal-all}. 
	
	The following features are apparent in Fig. \ref{fig:HorizonsCausal-all-2}:
	\begin{itemize}
		\item  Region 1: after formation the outer horizon (red) is spacelike and becomes null, and the inner horizon becomes timelike.
		
		\item Region 2: the density peak (green) crosses from outside the inner horizon to the region between the two horizons, the inner horizon is timelike, the density peak is timelike before crossing into the region between the two horizons and becomes spacelike after moving into that region.  
		\item Region 3:  the outer horizon is null for nearly all of its finite lifetime.
		
		\item Region 4 is the bounce phase: the inner horizon and density peak change from ingoing to outgoing. The inner horizon changes from timelike to spacelike, and the density peak (green) remains spacelike while inside the regions between the two horizons. The region to the left of the inner horizon is nearly flat. 
		\item Region 5: this is a near reflection of region 2. The density peak emerges from within the inner horizon and becomes timelike, while   the inner horizon remains spacelike.
		\item Region 6: the inner and outer horizons merge and the black hole region disappears. The peak matter density (not shown) is timelike and propagates outward. 
	\end{itemize}
	
	\noindent Throughout this process the peak matter density remains timelike while it is outside the region between the two horizons. The density profile is, however, not a thin shell and so has tails that straddle  both sides of the inner horizon while it's peak enters and then exits the region between the two horizons. It would require a more microscopic view to take into account the thickness of the density profile.  
	
	The details of the evolution of the inner horizon depends on the sharpness of the density profile derived from the mass function; a slower increase of the mass function (larger $\alpha$ in (\ref{m})) gives a wider density profile;  a linear combination of mass functions of the type (\ref{m}) with different values of $r_0$ is a monotonic function that rises from zero, remains constant for a radial range, and then rises again to its asymptotic value--this produces a  double peaked energy density with corresponding different behaviour of  apparent horizons.

\begin{figure}
		\begin{centering}
			\includegraphics[width=\columnwidth]{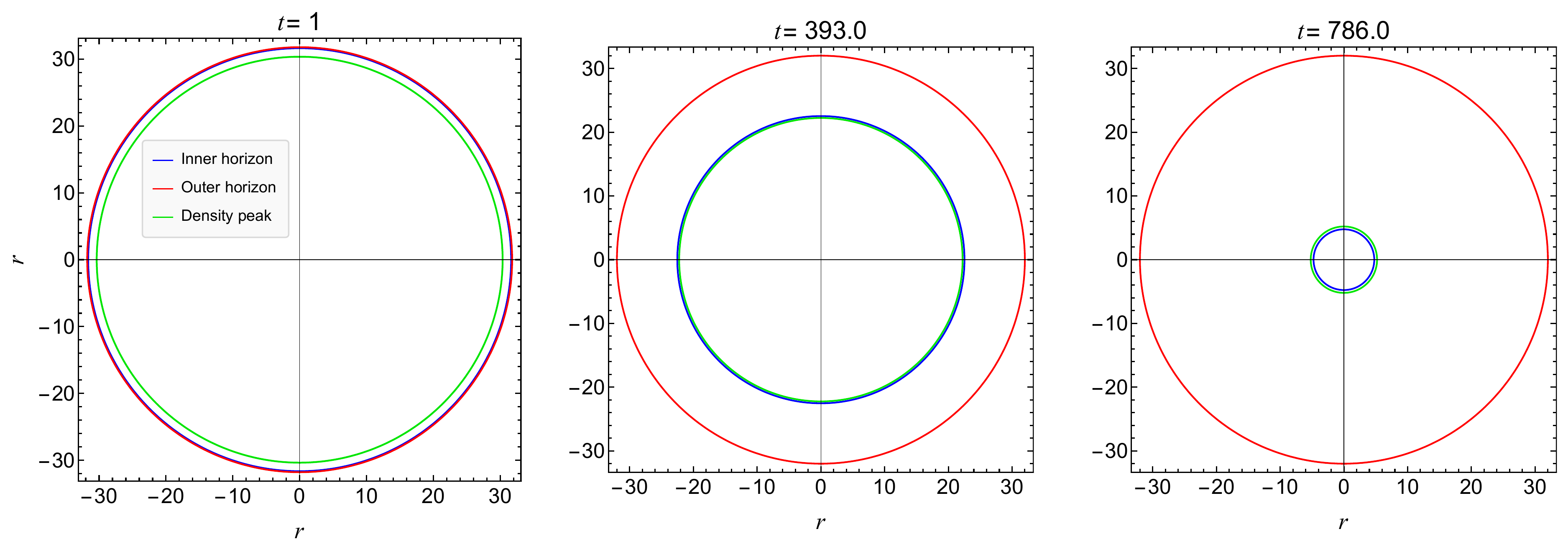}
			\par\end{centering}
		\caption{Radial views of horizons and peak matter density. Left: inner and outer horizons are nearly coincident with matter density peak at a smaller radius. Middle: inner horizon has moved inward and the density peak lies at a smaller radius just outside the trapped region between the horizons. Right: density density peak enters the trapped region  between the two horizons.} 
		\label{fig:PolarHorizonsAll}
	\end{figure}

\begin{figure}
		\begin{centering}
			\includegraphics[width=0.3\columnwidth]{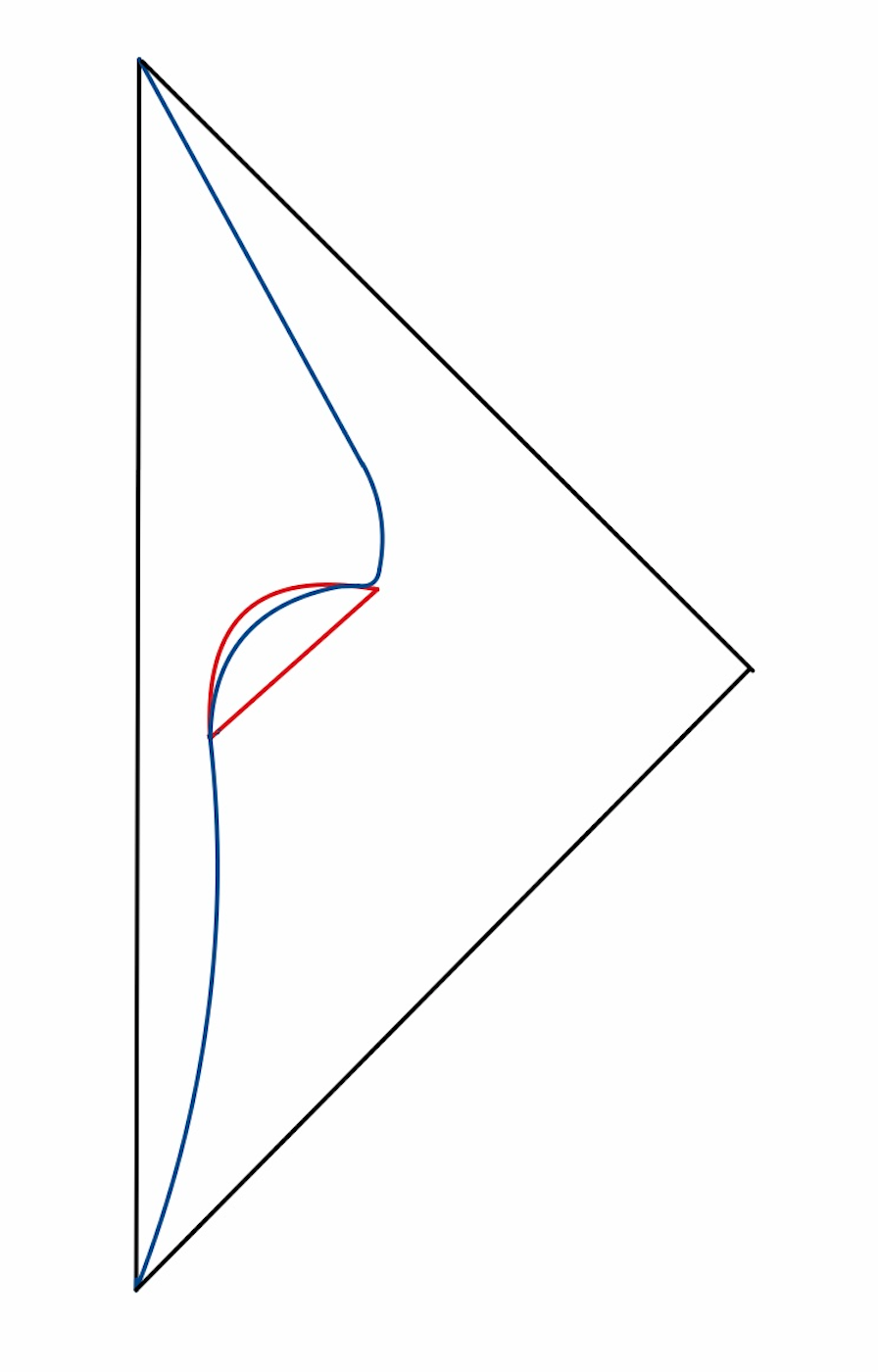}%
			\begin{picture}(0,0)
				\put(-15,110){$i^0$}
				\put(-145,110){$r=0$}
				\put(-60,160){${\cal J}^+$}
				\put(-60,62){${\cal J}^-$}
			\end{picture}
			\par\end{centering}
		\caption{Conformal diagram for the metric with mass function (\ref{m}). The red lines show the horizons. The outer horizon segment is null. The inner horizon is timelike before the bounce and becomes spacelike after the bounce. The blue line is the peak of the density profile. It is timelike outside the bubble region (between the horizons) and spacelike inside it.} 
		\label{fig:PenroseD}
	\end{figure}

	Fig. \ref{fig:PolarHorizonsAll} shows a radial view of the horizons and peak matter density for the same parameters: the first frame shows the configuration at $t=1$ where the inner and outer horizons are nearly coincident and the peak matter density is inside both horizons. The middle frame shows the inward movement of the inner horizon, and the last frame shows the situation near the bounce point where the peak matter density lies between the two horizons. 
	
	Fig. \ref{fig:PenroseD} shows the Penrose diagram deduced from Fig. \ref{fig:HorizonsCausal-all}. The red bubble is the region between the two horizons. The outer horizon is a null segment of what classically would be the event horizon. The inner horizon is timelike, then becomes spacelike after the bounce. The blue matter density peak follows the inner horizon, enters the bubble region, and then emerges before the horizons merge and disappear. Trapped surfaces are within the closed red region in the diagram, and are effectively shielded from infinity by the null outer horizon, i.e. there are no Cauchy surfaces that intersect the outer horizon and do not intersect the inner horizon.   

	\noindent 

	\subsection{Black hole lifetime}
	
	We define the black hole lifetime $T_{BH}$ as the time interval between horizon pair formation $t_i$ (when roots of $\Theta(r,t)$ first appear) and horizon annihilation $t_f$ (when roots of $\Theta(r,t)$ disappear):
	\begin{equation}
		T_\mathrm{BH}= t_f - t_i.
	\end{equation}
	For the case shown in Fig. \ref{fig:HorizonsCausal-all}, this is approximately the  time interval between regions 1 and 6. With a given velocity function of the form (\ref{v}), and Arnowitt-Deser-Misner (ADM) mass associated to the  mass function (\ref{m}),  $T_{\rm BH}$ is readily determined by numerically tracking the roots of $\Theta(r,t)$. (This process was also followed in \cite{Husain:2021ojz}, with the same definition of black hole lifetime.)
	
\begin{figure}
		\noindent \begin{centering}
			\includegraphics[scale=0.8]{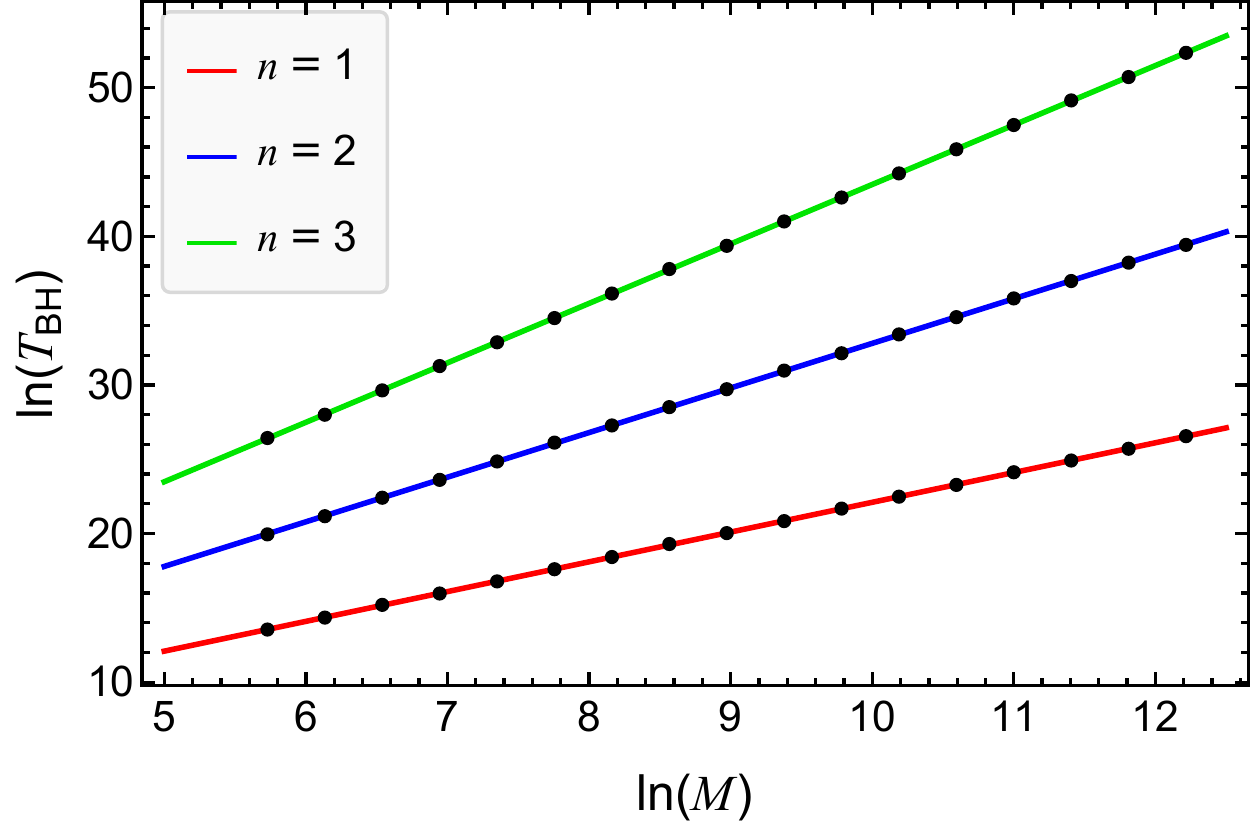}
			\par\end{centering}
		\caption{Black hole lifetime $T_{\rm BH}$ as a function of ADM mass $M$ with mass function (\ref{m}) and velocity  (\ref{v}). The lines are fit by the formula $T_{\rm BH}\approx 2^{n+2} M^{n+1}$} \label{fig:lifetime-compare}
	\end{figure}

	The power $n$ in the velocity function (\ref{v}) determines the ingoing and outgoing radially dependent velocity for a fixed parameter $A$; the larger the value of $n$, the slower the speed, and hence the longer the lifetime. This expectation is borne out: Fig. \ref{fig:lifetime-compare} shows log-log graphs of $T_{\rm BH}$ vs. ADM mass $M$ for $n=1,\ldots, 3$ with $A=1$ in (\ref{v}). The fit to these points is well approximated by the formula
	\begin{equation}
		T_{\rm BH}\approx2^{n+2}M^{n+1}.
	\end{equation}
	This expression quantitatively ties the power $n$ in the velocity function to the black hole lifetime as defined above. It contains the $M^2$ case calculated for dust collapse in \cite{Husain:2021ojz} (albeit without an outgoing shock wave since the metrics we are considering are smooth), and the Hawking evaporation case with lifetime proportional to $M^3$.
	
	\section{Conclusions and discussion}
	
	We have exhibited a non-singular spherically symmetric asymptotically flat time-dependent metric in PG coordinates that describes matter collapse and bounce of the type that might arise from a quantum theory of gravity. The main intuition built into the metric is that of a matter bounce through a smooth time-dependent mass function. 
	
	For early times before the bounce  (in the PG coordinate time), the metric could be that of classically collapsing matter in GR, i.e., outside the matter region it is Schwarzschild spacetime, and in the matter region it is a metric that satisfies the dominant energy condition. However, during and just after the bounce phase, the metric would correspond to a GR solution with matter that violates the dominant, null and strong energy conditions, but where the energy density $\rho$ in  \eqref{rho} remains positive and finite everywhere for all $t$; the spacetime region where the energy conditions are violated  first occurs in region 4 of Fig. \ref{fig:HorizonsCausal-all}, which coincides with the bounce region.  Horizon formation  and evaporation as shown in Figs. \ref{fig:PenroseD} follow directly from the features built into the mass function.  
	
	There are other physical quantities that can be calculated for the type of dynamical metrics we have considered. One such quantity is surface gravity. This has several useful definitions (see e.g. \cite{Vanzo:2011wq} for a summary); one such is due to Hayward \cite{Hayward:1993wb}, which for the metric \eqref{eq:Gen-metric} gives $\displaystyle \kappa =   (1+2\dot{m}-2m')/4m$, with dot and prime denoting derivatives with respect to $t$ and $r$ respectively. On the null portion of the outer horizon in our case, this quantity becomes 1/4$M$, with $M$ defined as in \eqref{ADM}, since all matter is within this outer horizon for its duration.

	Whether such a metric arises as an effective description from a quantum theory of gravity is an open question. However, quantum resolution of the singularity rules out the standard classical collapse and Hawking's ``information loss'' conformal diagrams, as both of these contain the $r=0$ spacelike singularity inside the event horizon. (Indeed, there is no known metric which gives the information-loss diagram due to the corner where the $r=0$ singularity transitions to a timelike curve). In the metric we propose, the classical event horizon is replaced by a segment of it, along with an inner-horizon, both of which last for a finite duration depending on the bounce velocity function.          
	
	There are other conjectured proposals for nonsingular black holes with corresponding conformal diagrams. These include \cite{Ashtekar:2005cj,Hayward:2005gi} which are similar in spirit to what is presented here, but without an explicit metric. Ref.  \cite{Martin-Dussaud:2019wqc} on the other hand describes a black hole to white hole transition, where the black hole event horizon connects directly to a white hole horizon, but again no explicit non-vacuum metric is presently available. The conformal diagram  derived from quantum gravity corrected dust collapse in \cite{Husain:2021ojz,Husain:2022gwp} is similar to Fig. \ref{fig:PenroseD}, but with a shock wave that arises in that calculation. 
	
	Lastly, there is a possibility not envisioned in the scenarios just summarized, and not contained in the metric we have presented. This corresponds to the situation where matter collapses,  bounces and expands outward, and then eventually begins to recollapse. The process then repeats ad-infinitum. This seems physically reasonable since classical gravity should be valid in regions of low curvature. It could be modeled by an oscillating velocity profile, different from the one we give in (\ref{v}). In the final analysis, it is a computational question for a quantum theory of gravity  which of these scenarios accompany singularity avoidance. There are other possibilities such as a quasistatic hovering ball of matter where the inward collapse tendency is balanced by the outward singularity avoiding ``pressure.''     
	
	\begin{acknowledgements}
		We thank Saurya Das for discussion and Edward Wilson-Ewing for comments on the manuscript. This work was supported in part by the Natural Science and Engineering Research Council of Canada. S. R. acknowledges the support of the Natural Science and Engineering Research Council of Canada, funding reference No. RGPIN-2021-03644 and No. DGECR-2021-00302.
	\end{acknowledgements}
	
	\appendix
	\bibliographystyle{apsrev4-2}
	\bibliography{mainbib}
	
\end{document}